\documentclass{aidb}
\usepackage{graphicx}
\usepackage{subcaption}
\usepackage{balance}  
\usepackage{color}
\usepackage{url}
\usepackage{multirow}
\usepackage{xspace}
\usepackage{float}
\vldbTitle{It's AI Match: A Two-Step Approach for Schema Matching Using Embedding (Regular Paper)}
\vldbAuthors{Benjamin Hättasch, Michael Truong-Ngoc, Andreas Schmidt and Carsten Binnig}

\begin{document}

\newcommand{\smiff}[1]{}
\newcommand{\naive}{na\"{\i}ve\xspace}
\newcommand{\Naive}{Na\"{\i}ve\xspace}
\newcommand{\naively}{na\"{\i}vely\xspace}

\widowpenalty = 10000


\title{It's AI Match: A Two-Step Approach for Schema Matching Using Embeddings (Regular Paper)}



%
%
%
%

\numberofauthors{4}

\author{
%
%
\alignauthor
Benjamin Hättasch\\
       \affaddr{TU Darmstadt}
\alignauthor Michael Truong-Ngoc\\
       \affaddr{TU Darmstadt}
\alignauthor Andreas Schmidt\\
       \affaddr{KIT \& Hochschule Karlsruhe}
\and  
\alignauthor Carsten Binnig\\
       \affaddr{TU Darmstadt}
}
\additionalauthors{}
\date{September 2020}

\maketitle

\begin{abstract}
Since data is often stored in different sources, it needs to be integrated to gather a global view that is required in order to create value and derive knowledge from it. 
A critical step in data integration is schema matching which aims to find semantic correspondences between elements of two schemata.
In order to reduce the manual effort involved in schema matching, many solutions for the automatic determination of schema correspondences have already been developed.

In this paper, we propose a novel end-to-end approach for schema matching based on neural embeddings. 
The main idea is to use a two-step approach consisting of a table matching step followed by an attribute matching step.
In both steps we use embeddings on different levels either representing the whole table or single attributes.
Our results show that our approach is able to determine correspondences in a robust and reliable way and compared to traditional schema matching approaches can find non-trivial correspondences.
\end{abstract}

\section{Introduction}

\paragraph*{Motivation} The quality of decisions is directly influenced by the available data and the ease of access to it. 
Important decisions should thus be made based on a holistic view using all available data. 
However, data is often scattered into different heterogeneous sources.
This is not only the case for many data science projects that need to integrate data from different independent sources but is also relevant within companies, where data typically resides in different systems. 
Data Integration can help to mitigate these issues since it allows to create a global view over independent data sources.

A critical step in data integration is schema matching which aims to find semantic correspondences between elements of two schemata.
Traditionally, this was done by experts with a good understanding of the semantics of the data \cite{hull1997managing}.
However, modern schemata are becoming larger and more complex, making manual schema matching both more time-consuming and more error-prone \cite{rahmbern}.
In order to reduce the manual effort involved in schema matching, many solutions for the automatic determination of schema correspondences have already been developed \cite{do2002coma, DBLP:conf/esws/PinkelBJMRSSK15, chen2012node, george2005understanding, ulfleser, mehdi2017approach, nozakietal, partyka2009semantic, sahay2019schema, rahmbern}.

\paragraph*{Contribution} A major problem of many automatic schema matching approaches is that they fail if the semantic similarity is hard to detect.
For example, instance-based column matchers typically fail to match columns that contain disjoint but semantically similar values such two tables with different street names or even worse the same content in different languages (e.g., French and English).
Another example are name-based matchers that rely on sources such as \textit{WordNet} to identify column matches:
while these approaches can detect hard-to-match cases (such as columns that use synonyms as names), they fail if this knowledge is not encoded in the resource.

In this paper, we thus propose a novel end-to-end approach for schema matching based on neural embeddings to mitigate these issues. 
The main idea is to use a two-step approach consisting of a table matching step followed by an attribute matching step.
In both steps we use embeddings on different levels (i.e., representing the whole table but also only single attributes).
This allows our approach to find non-trivial correspondences such as those discussed before.

Summarized the main contributions of this paper are:\\
(1) First, we provide an analysis of existing approaches to automatically support schema matching. 
(2) We then present our end-to-end approach for schema matching using neural embeddings. 
(3) We propose and analyze different matchers on multiple levels (i.e., tables and columns) to identify a set of possible table and attribute correspondences. 
(4) We evaluate our approach on several benchmarks and real-world data sets. Our results show that our approach is able to determine correspondences in a robust and reliable way and compared to traditional schema matching approaches can find non-trivial correspondences.

\paragraph*{Outline} This paper is structured as follows:
In Section \ref{sec:related} we give an overview about existing approaches for schema matching.
Afterwards, in Section \ref{sec:embeddings} we show how neural embeddings can be used for schema matching in general and which challenges and problems can arise.
In Section \ref{sec:approach} we give an overview of our two-step approach, which will be further elaborated and evaluated in Sections \ref{sec:table_matching} and \ref{sec:attribute_matching}.
Finally, in Section \ref{sec:conclusion}, we summarize our results and give an outlook on possible further work.

\section{Previous Approaches}
\label{sec:related}

Conventional supporting approaches for schema matching use syntactical features like \textit{Levenshtein distance} or \textit{n-grams} to compute the similarity between different schema information. 
More recent approaches also try to consider semantic aspects such as synonyms and hypernyms or rely on machine learning (ML).
In the following, we give an overview of the existing matching approaches and then discuss their shortcomings compared to our approach. 

There already exist multiple matching frameworks that integrate many of the before mentioned matchers or even combine them in so called hybrid matchers.
One prominent example is the COMA/COMA++ framework by Do and Rahm that provides reference implementations for various approaches \cite{do2002coma}.
Due to space constraints, however, we are not discussing the details of these matching frameworks in this paper.

\subsection{Schema-Based Approaches}

Schema-based approaches, as the name already implies, try to derive similarities from the available schema information such as tables and attribute identifiers, available comments and data types as well as constraints.

Islam and Inkpen \cite{islam2008semantic} present a method which derives the similarity of two texts by a combination of semantic and syntactic information. 
Their approach initially determines syntactic similarity using a modified version of the Longest Common Subsequence (LCS) metric, which provides an alternative method for the edit distance. 
Subsequently, the semantic similarity is determined corpus-based. 
They also propose an optional word order similarity calculation, which classifies two texts as similar if common words have the same order in their respective text. 
The final result for the similarity of two texts is then derived by normalizing and combining the three similarity metrics. 
This method has the disadvantage that a significantly large corpus of schema similarities must be available in order to learn enough semantic knowledge.

The approach of Lee et al. \cite{lee2009reconciling} compares the semantics of attributes to overcome semantic heterogeneity in the healthcare domain. 
It consists of two components: word similarity and word affinity. 
The similarity of two words is determined using domain-specific knowledge and \textit{WordNet} (synsets). 
The affinity of two words is deduced from overlapping characters, hyponymy and hyperonymy relationships. 

Chen et al. \cite{chen2012node} present a hybrid algorithm that determines correspondences using concepts. 
To extract the concepts, again \textit{WordNet}-synsets as well as the tree structure of the hyperonymy-hyponymy relationship is used. 
Afterwards, concepts are compared both syntactically (name) and semantically (WordNet node) and conceptual relationships at node level are transformed into conjunctive normal forms to confirm semantic relationships and derive similarity values.

\subsection{Instance-Based Approaches}

In contrast to schema-based approaches, instance-based approaches attempt to derive similarities between attributes from the values belonging to them.

Partyka et al. \cite{partyka2009semantic} examine the values of the attributes to be compared and try to derive their similarity to the \textit{entropy-based distribution} (EBD).

While earlier approaches determined this distribution by means of n-grams, which are, however, highly dependent on overlapping instances, Partyka et al. take a different approach to determine the EBD. Their \textit{TSim} algorithm uses the \textit{normalized Google Distance} (NGD), together with the cluster method \textit{K-Medoid}: 
Individual keywords are extracted from the instances and afterwards grouped by K-Medoid. 
EBD values for these clusters are then calculated by semantic comparisons of the instances with the NGD to derive attribute correspondences.

Mahdi and Tiun \cite{mahdi2014utilizing} combine strengths of regular expressions for numerical instances and semantic relations from \textit{WordNet} for text-based instances to determine correspondences instance-based. 
Similarly, Mehdi et al. \cite{mehdi2017approach} propose a combination of regular expressions and an external auxiliary source. 
Attributes are first classified as numeric, text-based, or mixed based on their instances. 
Samples of instances from each class are then extracted and compared. Regular expressions are applied to numeric and mixed instances to examine instances for syntactic similarities based on patterns. 
To measure the semantic similarity of text-based instances the Google Similarity Distance (\textit{Google Similarity Distance}) \cite{cilibrasi2007google} is used.

\subsection{ML-Based Approaches}

\textit{WordNet} and other lexical dictionaries are limited in their scaling and do not cover all terms. 
For this reason, Zhang et al. \cite{zhang2014ontology} present a name-based ontology matching approach based on word embeddings (vector representations of words, see next Section for more details) of the \textit{Word2Vec} model. 
A hybrid approach combining word embeddings and editing distances achieved the best results in their experiments.

The combined approach of Fernandez et al. \cite{fernandez2018seeping} also relies on syntactic comparison methods and \textit{Word2Vec}. 
They additionally present the concept of \textit{Coherent Groups} for combining word embeddings for compound words. 
Their approach first determines correspondences using syntactic comparison methods such as the edit distance for attribute names and the Jaccard similarity for instances. 
Word embeddings for names are then used to both remove incorrect correspondences and add missing ones. 
In a comparison between the pre-trained \textit{Word2Vec} and a domain-specific trained model, the former achieved better results.

The approach of Nozaki et al. \cite{nozakietal} is based purely on the word embeddings of \textit{Word2Vec}, which are used to semantically compare text-based columns and to determine correspondences with sufficiently high semantic similarity. 
For this purpose, word vectors are generated for all instances of a column; instances consisting of compound words by the sum of the word vectors of their subwords, with subwords being ignored if they do not appear in the dictionary.
The sum of all instance vectors finally forms the attribute vector.

Gentile et al. \cite{gentile2017entity} present a method to reduce the search space for entity matching. 
The approach attempts to group similar tables by representing tables with \textit{Word2Vec} vectors. 
They test three different approaches to generate a vector representation (table embedding) for a whole table:
attribute model, where the embedding is generated using the table attributes, entity model, which is based on the so called "subject column" (a column of type string with the highest number of different values), and a combination of both.
Their evaluation shows that table embeddings are a promising method for reducing the search space without explicit expert knowledge.

Chen et al. \cite{chen2019learning} designed a hybrid neural network that is trained to predict the semantics of table columns without metadata and only with instances. 
Table contents are vectorized using word embeddings and fed into the neural network as input for training and prediction.
In comparison to \textit{Word2Vec}, their RNN-based embeddings could provide better results.

Sahay et al. \cite{sahay2019schema} present a hybrid method that combines schema information and instance information to find similar attributes. 
A vector representation is generated using schema and instance features of a column. 
These vectors are then grouped using \textit{k-means Clustering} and only columns within clusters are compared using syntactic comparison methods such as edit distance.

\subsection{Discussion}

This short summarization shows that research in the field of automatic schema matching already produced a lot of approaches.
Hence, as mentioned before, simple schema-based or instance-based matchers have problems to discover semantic correspondences.
More recent approaches based on machine-learning aim to address these shortcomings similar to our approach; some of those approaches already propose to use neural embeddings. 
Different from those approaches, however, a major novelty of our approach is that we take a two-step approach that uses neural embeddings at different levels (i.e., tables and columns) and also to combine different aspects (i.e., schema information and instances).

\section{Using embeddings for Matching}
\label{sec:embeddings}

\begin{figure*}[t]
    \centering
    \includegraphics[width=\textwidth]{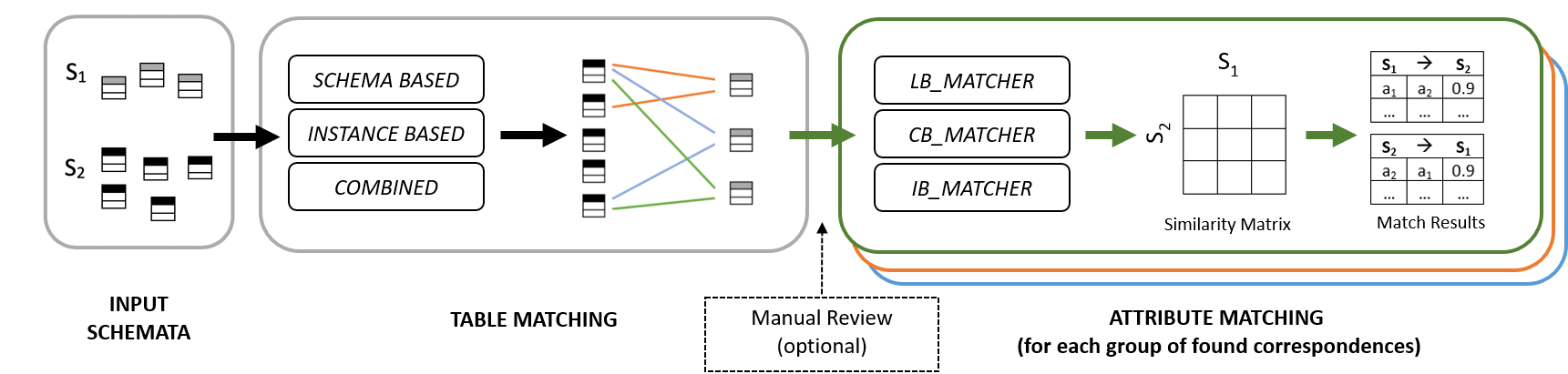}
    \caption{End-to-end process for schema matching, consisting of two steps (table matching and attribute matching).
    In the first step, possible matches between the tables are produced either on the basis of the available schema information (\texttt{SCHEMA BASED}), or on the basis of the instances (\texttt{INSTACE BASED}) of tables or using both together (\texttt{COMBINED}). 
    Optionally, after the first step a user can confirm or reject possibel table matches.
    For the next step, matches between the individual attributes of the remaining table pairs are determined. 
    This can be done using the table and attribute names (\texttt{NB\_MATCHER}), natural language comments in the database schema (\texttt{CB\_MATCHER}), or instances of a table (\texttt{IB\_MATCHER}).
    Either thresholding or ranking can be applied to the resulting similarity values to determine the final matches.
    }
    \label{fig:end_to_end}
\end{figure*}

In this section, we first will provide the necessary background to explain why neural embeddings (in particular word embeddings) provide a good starting point for schema matching.

\subsection{Similarity based on Embeddings}
\label{subsec:embedding_similarity}

In a nutshell, word embeddings represent words by vectors and map the---in many languages---unlimited number of words to a space with a fixed number of dimensions.
One usually speaks of embeddings if the representation, in contrast to, for example, one-hot-encoding, contains information that goes beyond syntactic features, i.e. the words are not arranged arbitrarily in vector space, but rather from a semantic point of view.
A typical possibility is to represent words by means of other words in whose context they often occur.

What makes word embeddings interesting for schema matching is that they can be used to compare semantic similarity between strings, hence improve over simple syntactic approaches like string-edit distances.
In the case of schema matching, for example, attribute names can be represented by word embeddings.
The similarity of two embeddings is computed by applying vector distance metrics, especially cosine similarity \cite{Mikolov:2013:DRW:2999792.2999959} which can be computed with low computing capacities using simple operations.
Although vector components of word vectors of different word embeddings models can be negative, word vectors are often trained in such a way that the cosine similarity is between 0 and 1 in almost all cases and negative only in very rare cases \cite{timbrueck}. 
Values near 0 indicate low semantic similarity, while values near 1 indicate high semantic similarity.

In contrast to simple string comparison, word embeddings can cover different syntactical variations for the same or similar concept (e.g., synonyms or abbreviations) and for contextualized word embeddings also different semantic meanings with the same syntax (i.e. ambiguities).
However, there are also additional challenges: 
one difficulty is the merging of semantics and relationship of words \cite{hill2015simlex}. 
Words like \textit{coffee} and \textit{cup} are semantically not identical, but are often mentioned in the same context.
This relationship is expressed in similar representations for many kinds of embeddings \cite{faruqui2016problems}.

\subsection{Different Embedding Approaches}
\label{subsec:embedding_approaches}

Word embeddings are usually trained on large corpora of text and therefore cover the relations between common words. 
The first generation of word embeddings like  \textit{Word2Vec} \cite{Mikolov:2013:DRW:2999792.2999959} or \textit{GloVe} \cite{pennington2014glove} were explicitly created for comparing single words. 
There exist simply to obtain pre-trained models that can be used for this task without additional knowledge about there internal function or need for training \cite{Kenter:2015:STS:2806416.2806475}. 
Although these models were trained on large corpora (e.g., the English Wikipedia or Google News), their vocabulary is still limited:
As a result, they do not support most of the multi-word expressions directly and it is not possible to generate word representations especially for domain specific terms.

There are several approaches to generate embeddings for expressions consisting of multiple words like sentences or paragraphs \cite{le2014distributed, socher2012semantic}. 
However, such models do not work very well for combining attribute names or instances because the combination of table values does not form grammatically correct sentences \cite{fernandez2018seeping}.

An alternative to off-the-shelf sentence embedding models is to use the average vector of all individual words with pre-trained embeddings for single words.
This method was developed by Castro Fernandez et al. \cite{fernandez2018seeping} using \textit{Word2Vec} for matching schema attribute names. 
They concluded that comparing the average vectors of multi-words is not an optimal solution for attribute matching and instead proposed \textit{Coherent Groups} as an alternative method for comparing multi-words with \textit{Word2Vec}.
Coherent Groups consider multi-word expressions as groups of individual words.
The method calculates the similarity of two groups of words, such as two attributes, by comparing the individual words in pairs. 
The average of all comparisons then determines the similarity between the two attributes.

Another problem are out-of-vocabulary (OOV) errors:
traditional word embedding models can only represent words they have already seen during training, which works well for texts about common topics but may fail for databases with domain-specific element names and content.
The amount of out-of-vocabulary words can be reduced by training or fine-tuning on domain specific texts. 
Yet this requires computing resources and suitable training corpora. 
The usage of embeddings only trained using the \textit{Distributional Hypothesis} \cite{harris1954distributional} (i.e. words are represented by the context they usually appear in) is therefore difficult for schema matching.

Contextualized word embeddings models like \textit{ELMo} \cite{peters2018deep}, \textit{BERT} \cite{devlin2018bert} or Google Universal Sentence Encoder (\textit{USE}) \cite{cer2018universal} are trained for additional tasks like machine translation (using both suitable data and an adequate internal model architecture), so that the resulting word vectors are semantically more meaningful. 
Since they can also generate word vectors dynamically (i.e. find a representation for unseen words), no methods for the aggregation of multiple words and handling of OOV are required.
This is done using WordPiece tokenization and breaking down unseen or rare words into subwords or characters.
Therefore partial semantic meanings (for segmentation into subwords) or at least syntactic information (for character level encoding) are preserved even for complex unknown terms and expressions and the downsides of the above mentioned approaches can be avoided.

During the pre-training of these models, attention was paid to covering as much semantic variability and many domains as possible. 
Therefore, the pre-trained models should be sufficient for many application purposes. 
If, however, many domain-specific acronyms are used in a database (especially in the schema information), a preprocessing step with a domain-specific dictionary can be helpful.
For example, it can be assumed that the generated word vector for \textit{purchase order} is semantically more meaningful than that for \textit{PO}.

For the evaluation of our approach in the next sections, we will mainly use a multi-lingual \textit{Google USE} model that was trained on 16 different languages.\!\footnote{\url{https://tfhub.dev/google/universal-sentence-encoder-multilingual-large/3}}
Yet, the described approach works with any contextualized word embedding.

\section{Overview of Our Approach}
\label{sec:approach}
In the following, we present our novel end-to-end process for schema matching. 
As mentioned before, our approach consists of two steps that build on each other (table matching and attribute matching). 
For each step, we will propose and test different matchers based on neural embeddings. 
An overview of our approach can be seen in Figure \ref{fig:end_to_end}.

While neural embeddings have been used already for individual tasks of schema matching (i.e., table or attribute matching), we suggest a new holistic approach that uses neural embeddings on different levels to combine table and attribute matching.
To be more precise, in a first step we use neural embeddings to match the elements of a schema on the table level. 
For each table in the target schema, we propose either all tables from the source schema where the similarity is above a certain threshold or the $n$ matches with the highest similarity.
In the second step, the system then determines suggestions as to which attributes from the corresponding tables fit together.
In addition, after the first step a user can optionally review and select those table pairs that should be kept for the second step.

\begin{figure*}
\centering
\begin{subfigure}[b]{0.42\linewidth}
\centering
\includegraphics[width=\textwidth]{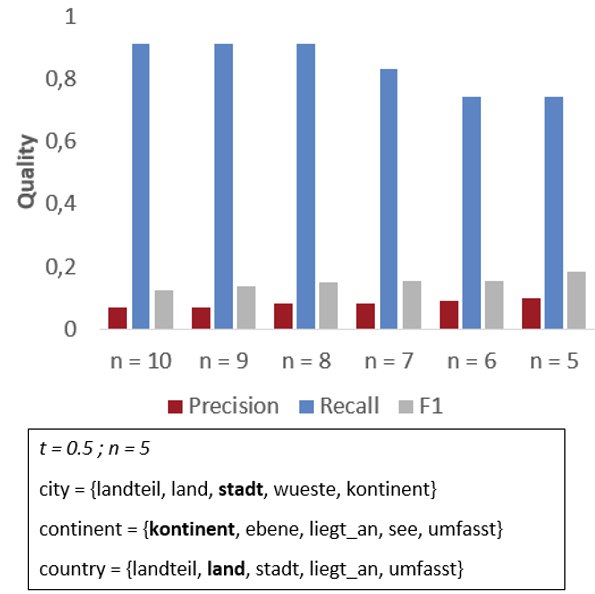} 
\caption{fixed $t=0.5$ and varying $n$ values}
\label{sfig:table_matching_n}
\end{subfigure}%
\hspace{8ex}
\begin{subfigure}[b]{0.37\linewidth}
\centering
\includegraphics[width=\textwidth]{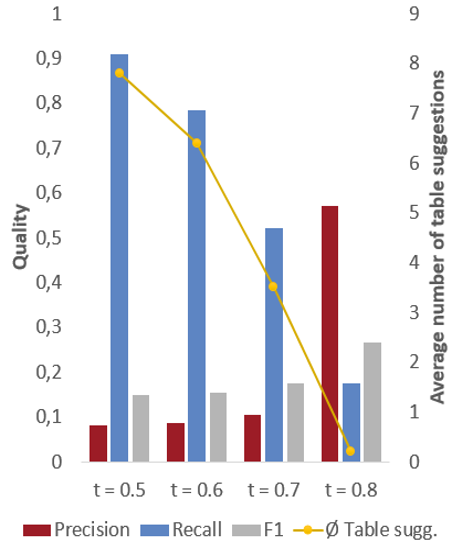}
\caption{fixed $n=8$ and varying $t$ values}
\label{sfig:table_matching_t}
\end{subfigure}%
\caption{Experiment 1 --  Cross-lingual schema-based table matching on geographic databases with different parameter settings. Precision, recall and F1 metric are reported. For a) a qualitative analysis of one setting is shown, for b) the average number of found possible correspondences for different similarity thresholds $t$ can be found.}
\end{figure*}

In the following, we give give an overview of the two steps of our matching procedure.

\paragraph*{Table Matching}
Two general approaches are possible here:
First, the schema information can be used in the form of table name and attribute names. 
Tables with semantically similar schema information are probably used to store similar content. 
Second, exactly these contents of the tables can be examined:
we can use the contents of each table to compute embeddings for the tables (either based on all attributes or a subset of it).
These embeddings can then be compared to other embeddings to find possible (partial) table matches.
Since the candidate pool for this comparison is a cross product of the attributes of all tables which we wanted to avoid with the two-step approach and the calculation of the data embeddings can be expensive to compute,
a combined approach is advisable: 
first the number of table pairs to be examined is reduced schema-based and then the candidate pool is further reduced using an instance-based approach.
For optimization, some intermediate results of the table matching can be stored for the attribute matching step following afterwards.

\paragraph*{Attribute Matching} 
In this step, we only use the table candidates that qualify based on table matching as discussed before.
For the attribute matching, we can again use schema or instance information analogous to table matching. 
If no further auxiliary information (e.g., strict type annotations) is available, all attributes of a source table must be compared with those of the target table. 
In this paper, we are considering again two different types of matchers: 
Name-based matching uses the table information such as table name and attribute names, as well as available comments. 
Instance-based matchers inspect the contents of the columns to find attributes with semantically equivalent instances.

\section{Step 1 -- Table Matching}
\label{sec:table_matching}

In the following, we discuss the details of our first step and present and evaluate different approaches for table matching using schema information and instance information for computing a table embedding.

\subsection{Structure-Based Matching}
\label{subsec:table_structure}

As discussed before, in a first step we use our table matcher to reduce the search space for the subsequent attribute matching step.
It should therefore recognize as many table correspondences as possible (while still delivering as few incorrect correspondences as possible to have a noticeable effect).
A high recall is hence more important than good precision.

Table matching is initially a ranking problem (ranking all target tables for a source table based on the supposed similarity). 
Afterwards, the pool of candidates has to be restricted: 
by taking the best $n$ correspondences, by applying a similarity threshold $t$, or both.
In our experimental evaluation, we therefore check if it possible to determine suitable values for these parameters.

Gromann et al. \cite{gromann2018comparing} have shown in their investigations that for aligning ontologies better results can be achieved with word embeddings than with traditional string comparison methods. 
We will now investigate, whether this also holds for matching relational databases.
Furthermore, we examine the use of next-generation embeddings that can dynamically generate vectors for each input, thus avoiding out-of-vocabulary errors and not requiring explicit handling of multi-word expressions.

For computing table similarity using a table embedding that relies on schema information, we represent each table by a vector for the (equally weighted) combination of table name and all attribute names. 
The representations are determined using the pre-trained Google USE model.

\paragraph*{Experiment 1: Schema-Based Table Matching}
In this experiment we test the matching exclusively on the basis of schema information.
The main goal of this experiment (as well as the other experiments in this section) is to show the matching quality of our approach for table matching in isolation.
An end-to-end evaluation and a comparison with existing approaches as baselines will be presented in the experiments in Section~\ref{subsec:attribute_instance}. 

For the table matching experiments, the English Mondial DB\footnote{\url{https://www.dbis.informatik.uni-goettingen.de/Mondial/}} and the German Terra DB\footnote{\url{https://www.sachsen.schule/~terra2014/index.php}} are used. Both databases contain geographic content such as mountains, rivers, deserts, or islands and have similar tables since the later one was derived from the first one for teaching purposes \cite{durr2013einsatz}.
We manually aligned them as a gold standard.
In the following, we show the results using these data sets with varying parameters. 

First, we test the effects of the parameter $n$ (i.e. only the best $n$ correspondences are considered) with a fixed minimum cosine similarity threshold $t$ of $0.5$.
The results can be seen in Figure \ref{sfig:table_matching_n}. 
75\% of the correspondences are detected with $n=5$, with $n>8$ even 90\% or more.
Second, we fix $n=8$ and vary the threshold $t$ (see Figure \ref{sfig:table_matching_t}). 
As expected, the average number of matching candidates shrinks for higher similarity thresholds, and so does the recall while the precision increases. 
At a threshold of $t=0.8$ only very few correspondences ($0.2$ per table on average) are suggested, hence the recall is low. 
However, the remaining suggestions have a good quality, which manifests itself in a high precision ($ > 0.55$).

\begin{figure*}
    \centering
    \includegraphics[width=.8\linewidth]{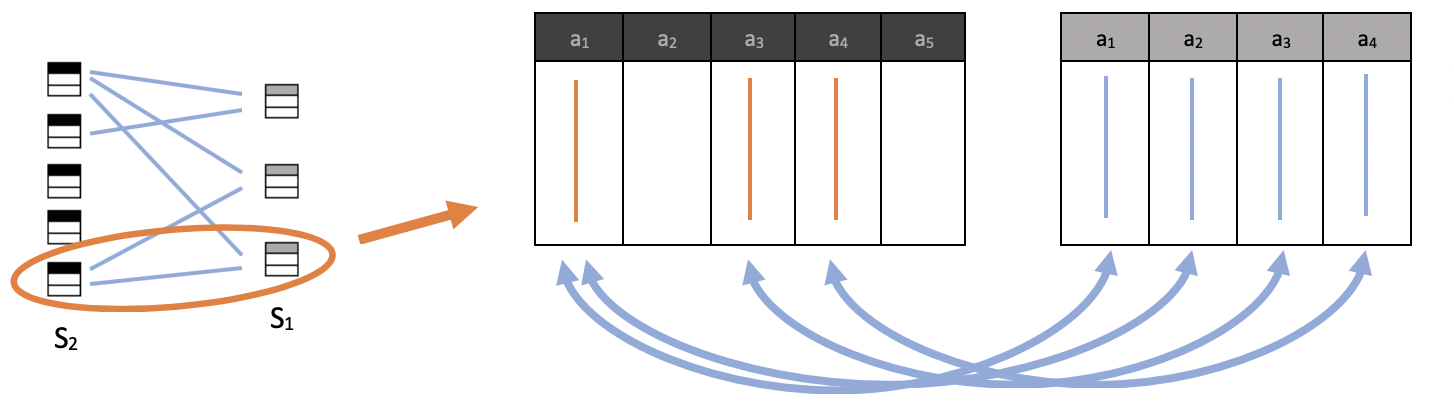}
    \caption{Instance-Based Table Matching -- The similarity of tables is inferred from the averaged similarity value for each attribute correspondence.}
    \label{fig:table_matching_instance_theory}
\end{figure*}

A qualitative analysis of the results shows that the reason for low precision values at lower thresholds is probably due to similar schematic information in the different tables.
For example, almost all tables contain the attribute \textit{name} in slightly modified form.
Nevertheless, the experiment shows that if the attributes are sufficiently unique, it is possible to determine schema-based table correspondences with embeddings.

\subsection{Instance-Based Matching}
\label{subsec:table_instance}

Alternatively to only using schema information to compute embeddings, we can also use instances to compute table embeddings.
Even with well-existing schema information, additional instance information can  help to further increase matching accuracy \cite{dohai}.
Studies show that traditional methods can even achieve better results with instance-based matching than with name-based matching (depending on the quality of the instances) \cite{ulfleser}.

In instance-based matching, a distinction is made between horizontal and vertical matching \cite{ulfleser}. Vertical matchers compare column contents of individual attributes and infer attribute correspondences, while horizontal matchers attempt to identify duplicates (i.e. two or more representations of the same object) between two schemata.
There exists work by Ebraheem et al. \cite{ebraheem2018distributed} for the recognition of duplicates in schemata (\textit{Entity Resolution} or \textit{Record Linkage}) with word embeddings. 
They investigate both the use and adaptation of pre-trained models, and the training of new embeddings for this purpose.

We propose an approach for vertical schema matching that uses embeddings as shown in Figure \ref{fig:table_matching_instance_theory}:
embeddings in our approach are used to represent the entire content of a column by a single vector and then to compare them with each other.
Since this paper aims at the evaluation of (word) embeddings, we only consider string attributes here.
If the resulting vector representations for attributes are sufficiently similar, an attribute correspondence can be assumed.

The most \naive{} approach to construct this vector is to determine the vector representations of all instances through the pre-trained model and then to average these to obtain the representation of the attribute.
This is inspired by the \naive{} but solid \cite{tudnlp} baseline approach for sentence embeddings, where a sentence is represented by the average of the individual word vectors of all its words.

However, this approach assumes that all instances are equally important.
Analogous to stopwords in classical NLP approaches, there may be instances that contribute less to the semantic meaning, for example placeholders like \textit{not-in-universe}, \textit{unknown}, \textit{NONE} etc. 
Such instances are also called noise \cite{tfidftabular} and may dominate the representation if no countermeasures are taken.
The averaging approach also ignores the order of instances, but these are usually irrelevant to the semantic meaning of an attribute \cite{george2005understanding}.

The aim is therefore to combine instances to form a single vector representation for attributes that most closely reflects the semantics of the attribute, using frequency values but also relativizing them if necessary (to avoid statistical bias \cite{mugo2002sampling}). 
Sampling based on frequency values can be used for this purpose.
At the same time, however, the frequency of an instance is not a clear indicator of whether the information is relevant or not:
if not applied carefully, sampling might amplify the noise and thus de-emphasize the representation.
Three possible sampling methods are:

\begin{description}
\item \textbf{Distinct Sampling} 
Ignore duplicates when generating the combined representation. 
This might lead to information loss if the instances are not equally distributed.
\item \textbf{N-Random-Sampling} 
Take $n$ random instances. 
Each instance is sampled with the same probability, hence to get a random distribution over distinct instances, distinct sampling should be performed first. 
The subset resulting from the selection is in most cases balanced if the sample sets are large enough so that the column is well represented. 
The method is considered the safest way to counteract a statistical bias in the resulting subset \cite{jawale2012methods}.

N-Random-Sampling can also be used to validate the semantic representation of a column with word embeddings: 
for this purpose, at least two sample sets are taken from the column containing randomly selected instances. 
The comparison of the semantic representations of the two samples  should give a cosine similarity close to one.
We will use this technique to test our instance representations in Experiment 6.

\item \textbf{N-Most-Common Sampling (with distinct sampling)} 
Here the instances that occur most frequently are selected. 
To do this, first a ranking by frequency is created and then each of the $n$ most frequent (distinct) instances is taken to compute the common representation.
This method will discard rare values and ensure that even frequent placeholders will only be included once in the representation
Depending on the size of the tables, the calculation can be much more complex than the other two sampling methods.
Hence, this method is only beneficial if there are certain instances that occur considerably more frequently than others.
\end{description}

After a representation has been found for the individual attributes, all attributes of the possible source and target table must now be compared to each other.
In a basic approach, all possible correspondences with a certain quality are searched for, which requires two threshold values:
First, it must be determined from which similarity of the attribute representation one can assume that they match.
Second, a ratio is required to determine the minimum fraction of attributes that must match in order for the tables to be considered match candidates.

The choice of this parameter depends on the desired matching scenario:
If one only wants to match tables that contain mostly the same types of data, the parameter must be set relatively high.
If, on the other hand, tables are also to be found that contain basically the same entities but have different purposes and therefore mostly store different attributes for these entities, it should be close to $0$.
In order to be able to distinguish whether two tables really have little in common or whether one table is only more fine-grained than the other, the comparison must also be carried out in both directions.
Unless a weighting of the columns is known, all attributes of a table should be considered for that calculation.

An alternative approach to directly comparing individual attribute similarities is to average the similarity values of all pairs of attribute representations to obtain a single similarity value for each pair of tables.
These can then be used to derive the best table correspondence(s) for a given table.
It can be useful to incorporate not all similarity scores for attribute pairs into this common similarity score for table pairs but to ignore all attribute pairs with a score below a certain threshold.
This allows it to obtain meaningful values for all matching problems where partial matches of tables are also relevant (i.e., not only the best 1:1 correspondence is searched for).
Further fine-tuning (e.g., additional normalization with the ratio of selected to overall attributes of a table) may be needed depending on the schemata to match.

\begin{figure}
    \centering
    \includegraphics[width=\linewidth]{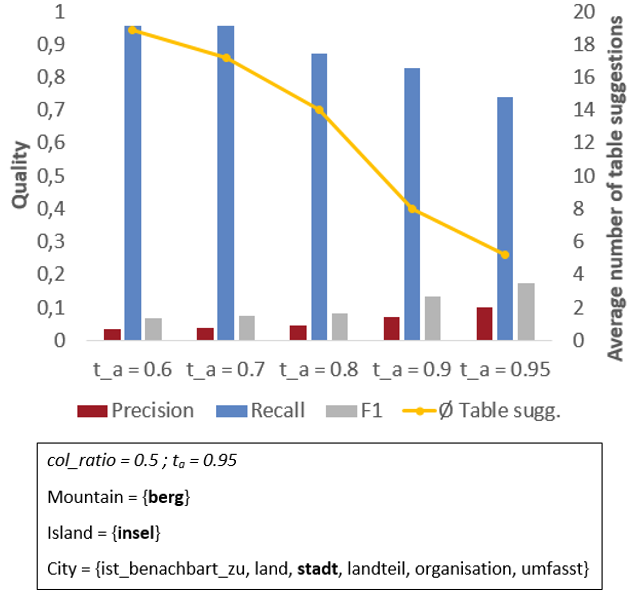}
    \caption{Experiment 2 -- Cross-lingual instance-based table matching on geographic databases. Precision, recall and F1 metric as well as the average number of found possible correspondences are reported for different attribute similarity thresholds $t_a$ and a fixed attribute match ratio of 0.5. We also show a qualitative analysis of a fixed parameter setting.}
    \label{fig:table_matching_instance}
\end{figure}

\paragraph*{Experiment 2: Instance-Based Table Matching}
In order to see the effect of the parameters of our instance-based table matcher, we again compare the two geographical databases.
Although the databases contain many similar instances, a syntactical matcher would find only a fraction of these matches, since one database contains English and the other German content.
We use N-Most-Common Sampling and only consider non-numerical attributes.
For each attribute in the target table, only the best match in the source table is considered (although it might be selected for multiple target attributes).
We require a match of at least half of the attributes ($col\_ratio = 0.5$) and test the effect of different thresholds $t_a$ above which two attributes are assumed to be similar.
The results can be found in Figure \ref{fig:table_matching_instance}.
With an instance-based threshold of $t_a = 0.95$, almost 75\% of the table correspondences can be found with an average of five table suggestions per target table.
A qualitative analysis shows that the representation of columns containing abbreviations or artificial IDs often resemble each other.
At this point, an additional syntactical comparison for exact matches could help to distinguish between them.

\subsection{Compound Grouping Approach}

For the instance-based matching described above, it would be necessary to compare the attributes of all tables with each other, hence the runtime will scale quadratically with the number of attributes and additionally depends on the number of instances.
It is therefore advisable to combine the two approaches and work on groups of tables (analogous to the fragment based matching by Do \cite{dohai}), i.e. to first reduce the number of tables to be compared by schema-based matching and then to exclude further candidates on an instance-based basis to increase precision.
To obtain a high recall, the thresholds must be selected as low as possible in the first step, whereby the selection should be based on the possible computing effort for the second step, since the computing effort for the first step is negligible in comparison.

\begin{figure}[t]
    \centering
    \includegraphics[width=\linewidth]{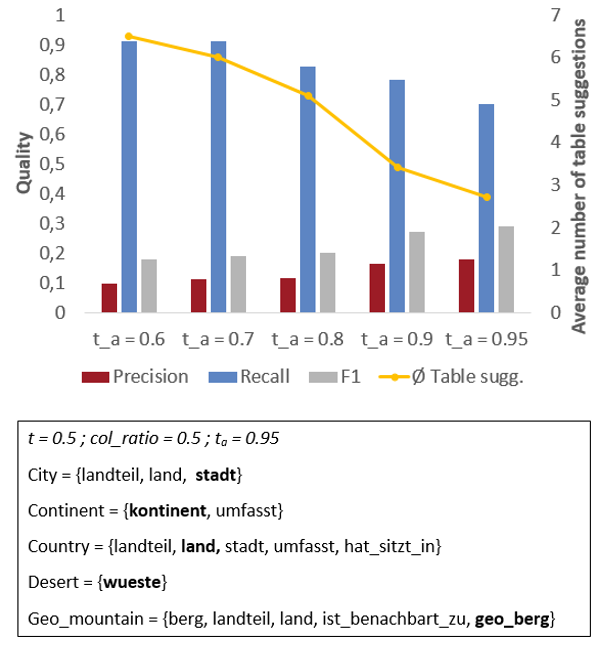}
    \caption{Experiment 3 -- Two-step cross-lingual table matching on geographic databases. In the first step, schema-based matching with $t=0.5$ is applied. Afterwards, instance-based matching is performed on the resulting groups. Precision, recall and F1 metric as well as the average number of found possible correspondences are reported for different attribute similarity thresholds $t_a$ and a fixed attribute match ratio of $0.5$. We also show a qualitative analysis of a fixed parameter setting.}
    \label{fig:table_matching_combined}
\end{figure}

\paragraph*{Experiment 3: Matching with Schema-Based Grouping}
In the final experiment for this section, a combined approach will be investigated.
For this purpose, all tables with a similarity of at least $t=0.5$ are determined as described in Experiment 1. 
These correspondences are then refined instance-based using the procedure described in Experiment 2, again using a $col\_ratio$ of $0.5$.
Figure \ref{fig:table_matching_combined} summarizes the results for different instance matching thresholds $t_a$. 
Compared to the experimental results of the individual strategies, the best F1 value was achieved with $t_a = 0.95$.
For each target table, an average of $2.7$ tables were suggested, with $70\%$ of the table correspondences being found.

The experiments show that the described procedure can be used to find correspondences that would not be recognized by syntactic matchers without, for example, requiring domain-specific ontologies. 
With the table matching step, it is possible to severely restrict the set of attributes to be compared, thus to considerably reduce the calculation time for attribute matching or to allow more complex operations to be performed per comparison.

\newpage
\section{Step 2 -- Attribute Matching}
\label{sec:attribute_matching}

In order to determine not only tables that might contain related content, but also exact correspondences between attributes of different tables, the candidates from the previous step must now be refined.
Depending on the parameter selection, either 1:1 relationships or a list of possible attribute correspondences are created, which can then be used directly or confirmed manually.
Table correspondences from table matching for which no attribute correspondence could be confirmed are automatically rejected.

Matching at attribute level is done in a similar way to matching at table level in many places, but the parameters such as weights and thresholds must be selected differently.
As before, both structural information (such as table and attribute titles or comments) and the actual instances can be used for this purpose.
In the following, the adjustments compared to table matching are explained and the individual approaches are evaluated.

\subsection{Name-Based Attribute Matching}

Section \ref{subsec:table_structure} already showed how table and attribute names can be represented and then compared using embeddings.
In this section, we use an embedding-based attribute-matcher which only considers the names of the individual attributes without including any additional information like neighboring elements or data type information.
The similarity of the attribute-matcher thus purely relies on the cosine-similarity of the embedding representation for two attribute names.

\begin{table}[]
    \renewcommand{\arraystretch}{1.5}
    \centering
    \begin{tabular}{lccc}
          & \textit{Precision} & \textit{Recall} & \textit{F1} \\
          \hline
    \textbf{COMA NodeNames} & 0.494 & 0.472 & 0.451 \\
    \textbf{Our Approach (NB)} & 0.502 & \textbf{0.516} & \textbf{0.507} \\
    \hline
    \textbf{COMA AllContextInst} & 0.577 & 0.465 & 0.489 \\
    \textbf{Our Approach  (IB)} & \textbf{0.979} & 0.257 & 0.405 \\
    \end{tabular}%
    \caption{Experiments 4 \& 8 --  Average values for name-based (NB) and instance-based (IB) version of our approach compared to the two COMA baselines. Values between 0 and 1, higher is better.}
    \label{tab:coma_use_avg}
\end{table}

\begin{figure*}
    \centering
    \includegraphics[width=\textwidth]{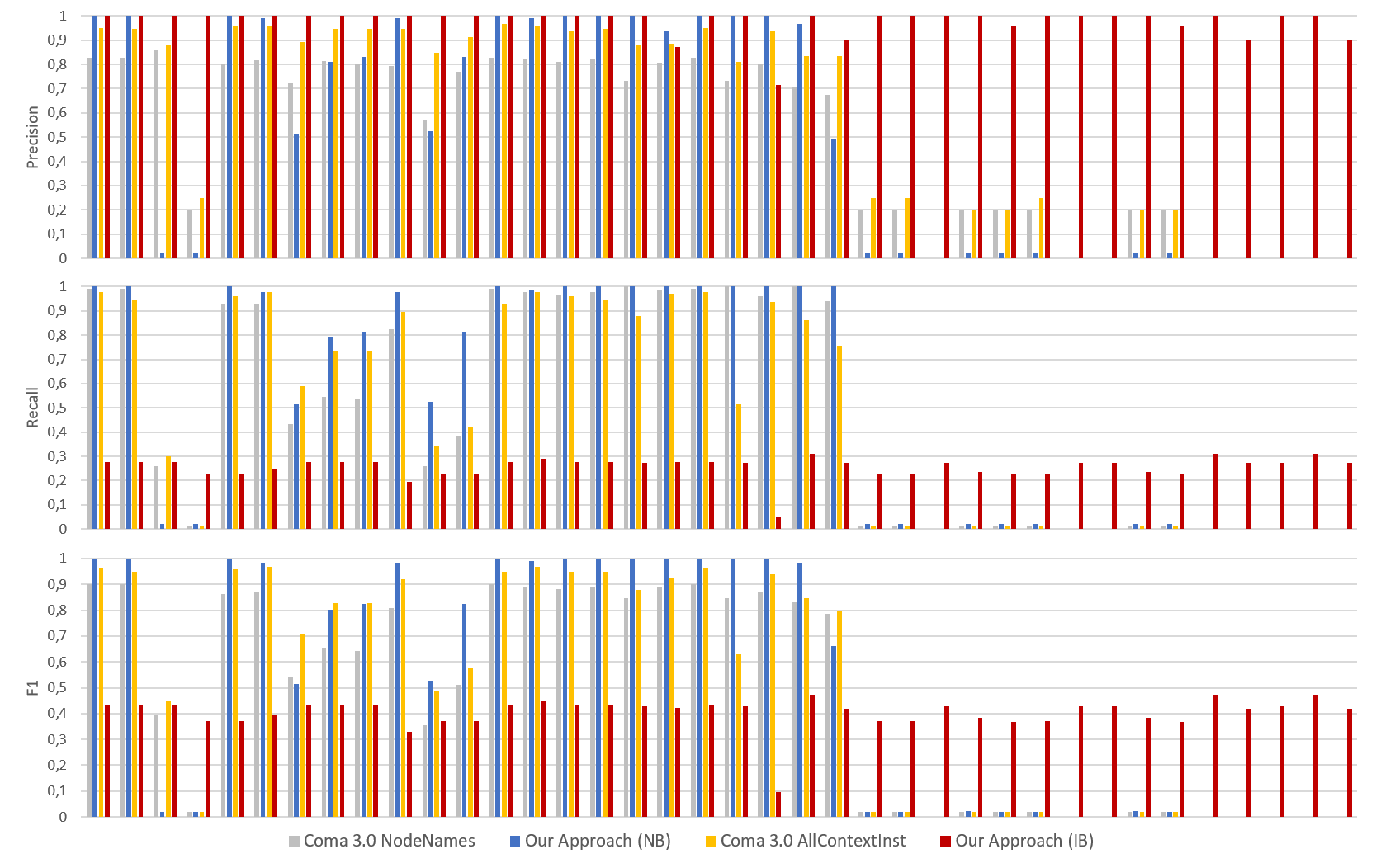}
    \vspace{-4.5ex}
    \caption{Experiments 4 \& 8 -- Precision, Recall and F1 Scores for applying the two COMA baselines and the name-based (NB) and instance-based (IB) variant of our approach to all 38 matching problems of the OAEI benchmark.}
    \vspace{-2.5ex}
    \label{fig:coma_38}
\end{figure*}

\paragraph*{Experiment 4: Name-Based Attribute Matching}
To compare the attribute-matcher with other matching baselines we use the OAEI Benchmark\footnote{Ontology Alignment Evaluation Initiative. \url{http://oaei.ontologymatching.org/2009/benchmarks/}} from 2009, which includes ontologies with different matching challenges. 
In this benchmark, a reference ontology comprising of 33 classes, 64 properties and 76 instances must be matched to modified ontologies. 
The modified ontologies embody 51 match problems, including, for example, the replacement of element labels with random or foreign words.
Since the data to be matched was created by modifying the individual attributes, there are 1:1 matches that must be found.

In order to enable comparability with other strategies, we will consider only the 38 matching problems in which there are instances in general.
However, the data is sparse, hence even for them only about 1/3 of the attributes contain instances.
As a baseline for comparison we use the COMA 3.0 framework already introduced in Section \ref{sec:related}, which provides reference implementations for standard matching techniques in so-called workflows.
These workflows usually consist of a combination of different (complex) matchers, i.e. the application of similarity metrics (partially using auxiliary information).

For this experiment, we use the  \texttt{NodesNameW} strategy of COMA 3.0, which only considers the attribute names and evaluates their similarities using syntactical comparison methods such as the Levenshtein distance to show how much a neural embedding matcher can improve over a standard matcher.
In later experiments, we also compare to more sophisticated matchers in COMA.
Different from the simple COMA matcher, our name-based matcher searches for the best correspondence between vector representations of attribute names.
It turns out that the embedding-based approach provides on average both better precision and higher recall (see Table \ref{tab:coma_use_avg}).
The results of the individual matching problems can be found in Figure \ref{fig:coma_38}.
It can be seen that our embedding-based approach almost always performs better for those problems where the name-based variant of COMA already performs well, but performs worse for problems where COMA already has difficulties.

\begin{table}[tb]
  \renewcommand{\arraystretch}{1.5}
  \centering
    \begin{tabular}{lcccc}
          & \textit{Aggr.} & \textit{Precision} & \textit{Recall} & \textit{F1} \\
          \hline
    \multirow{2}[0]{*}{\textbf{Our Appr. (W2V) }} & CG & \textbf{0.65} & 0.26  & 0.37 \\
           & Sum   & 0.35  & 0.61  & 0.44 \\
          \hline
    \multirow{2}[0]{*}{\textbf{Our Appr. (GloVe)}} & CG & 0.52  & 0.32  & 0.4 \\
           & Sum   & 0.27  & \textbf{0.69} & 0.39 \\
          \hline
    \textbf{Our Appr. (ELMo)} & -     & 0.43  & 0.58  & \textbf{0.5} \\
    \end{tabular}%
    \caption{Experiment 5 - Precision, Recall and F1 for matching on the XDR benchmark using static embeddings (Word2Vec (W2V) \& Glove) with Coherent Group (CG) or Sum Aggregation method compared to a contextualized embedding (ELMo). Values between 0 and 1, higher is better.}
  \label{tab:stat_vs_context}%
\end{table}%

\begin{figure*}[t]
    \centering
    \includegraphics[width=0.9\textwidth]{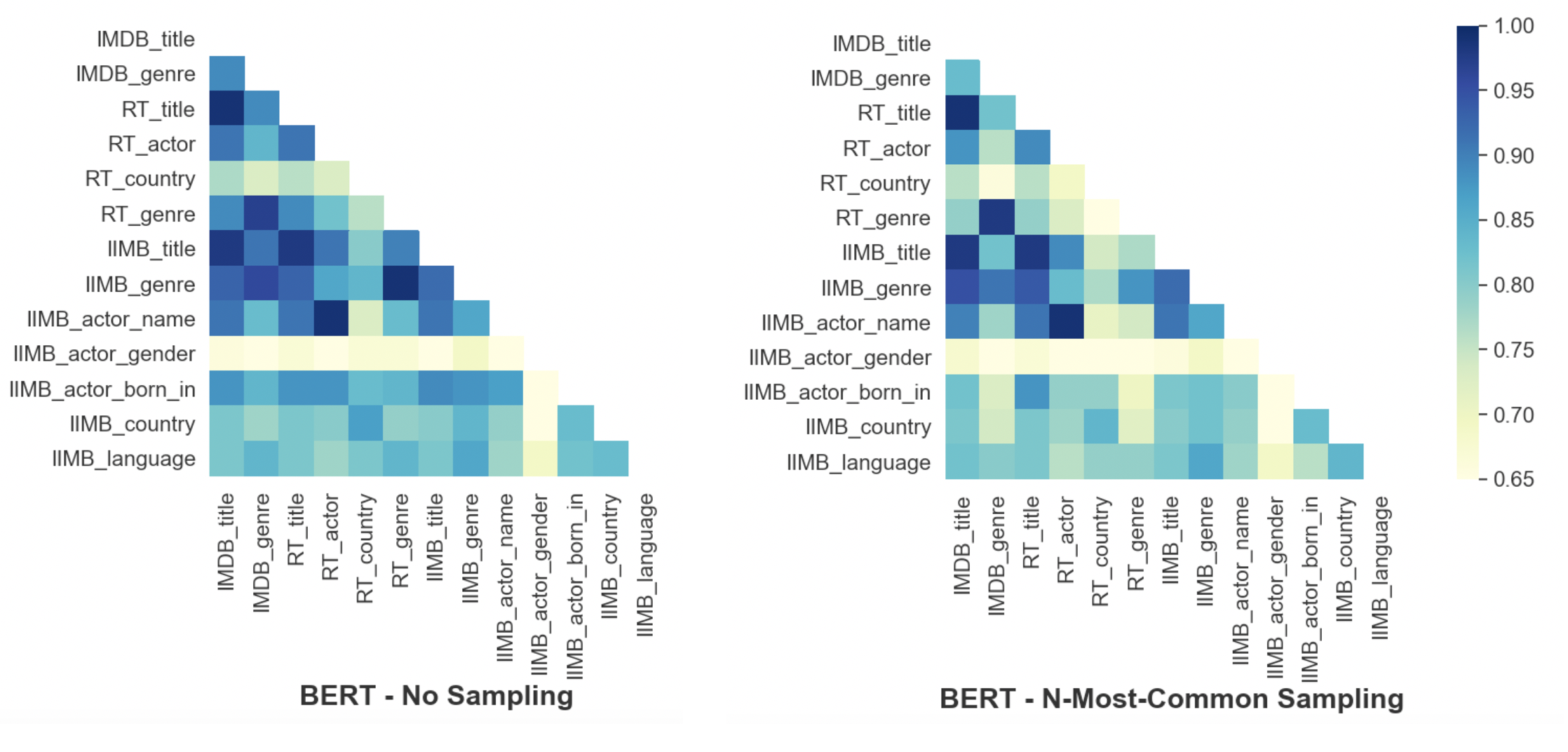}
    \vspace{-4.5ex}
    \caption{Experiment 7 -- Cosine similarities between instance representations for different movie attributes using BERT and different sampling approaches. Values between 0 and 1, similar attributes should have a value close to 1.}
    \label{fig:sample_effect}
\end{figure*}

\paragraph*{Experiment 5: Static vs. Contextualized Embeddings}

After having shown that using word embeddings has an advantage over traditional models, we will now examine the intuition mentioned in Section \ref{sec:embeddings} that contextualized embeddings are better suited for this purpose than traditional static embeddings.
The classical static embeddings \textit{word2vec} and \textit{GloVe} are compared with the contextualized \textit{ELMo}.
The latter takes the context into account, but in contrast to \textit{BERT} and \textit{Google USE} it is still unidirectional, allowing a fairer comparison.

For this experiment, we use the five XDR schemes\footnote{\url{https://dbs.uni-leipzig.de/bdschemamatching}} for customer orders (CIDX, Excel, Noris, Paragon and Apertum) as a benchmark. 
They do not contain instances, but only schema elements, and there are many abbreviations and composite element names such as \texttt{qty}, \texttt{contactName}, \texttt{POShipTo}, or \texttt{SupplierReferenceNo}.
In a pre-processing step, compound words were separated into partial words and known abbreviations such as \texttt{PO} were translated into their written form using a dictionary.

The average precision, recall and F1 score can be found in Table \ref{tab:stat_vs_context}.
For each possible schema pair, all attributes were compared using our representation approach with the different embedding models and all attribute pairs with a similarity of at least $0.8$ were selected as correspondence. 
For the Word2Vec and Glove models, a distinction was made between the two aggregation methods \textit{Coherent Groups} and \textit{Sum} (see Section \ref{sec:embeddings}).
For ELMo the representation could directly be computed using the model itself, regardless of whether the expressions were single or multiple words.

The two static embeddings \textit{word2vec} and \textit{GloVe} reach similar F1 values depending on the aggregation method. 
While the precision is higher with \textit{Coherent Groups}, more correspondences are found with the \textit{Sum} aggregation method. 
The contextualized model ELMo is a compromise between Precision and Recall and thus reaches the highest F1 value. 

The low recall values when using coherent groups can be explained by OOV errors.
For example, in the correspondence \texttt{(orderNum, customerOrderRef)} the subwords \texttt{num} and \texttt{ref} have no vectors and lead to a similarity of 0.
Furthermore, for Coherent Groups, the threshold of $0.8$ is very high, since the average value of all partial comparisons of the word groups tends to be lower due to unequal subwords. 
Therefore, in most cases only attribute correspondences with nearly or complete name equality were suggested.

Sum aggregation, on the other hand, produced many incorrect correspondences, since unequal attributes often contain identical subwords:
the attributes \texttt{contactName} and \texttt{companyName} incorrectly have a similarity of more than $0.8$, since half of the resulting word vectors consist of the vector for the subword \texttt{name}.

The contextualized word embeddings model ELMo also frequently produces incorrect correspondences due to identical subwords, but to a lesser extent than the sum aggregation method.
Since ELMo can form vectors for whole sentences, these vectors are not only generated purely from the vectors of the subwords, but also from the property of how subwords stand in context to each other.
For this reason additional synonyms like \texttt{(contactPhone, telephone)} or \texttt{(contactEmail, mail)} could be found.

The experiment demonstrates that it is reasonable to not perform the handling of multi-word expressions and domain-specific vocabulary through additional steps, but to use contextualized embeddings that can handle them inherently.

\subsection{Instance-Based Attribute Matching}
\label{subsec:attribute_instance}

\begin{figure*}
    \centering
    \includegraphics[width=.92\textwidth]{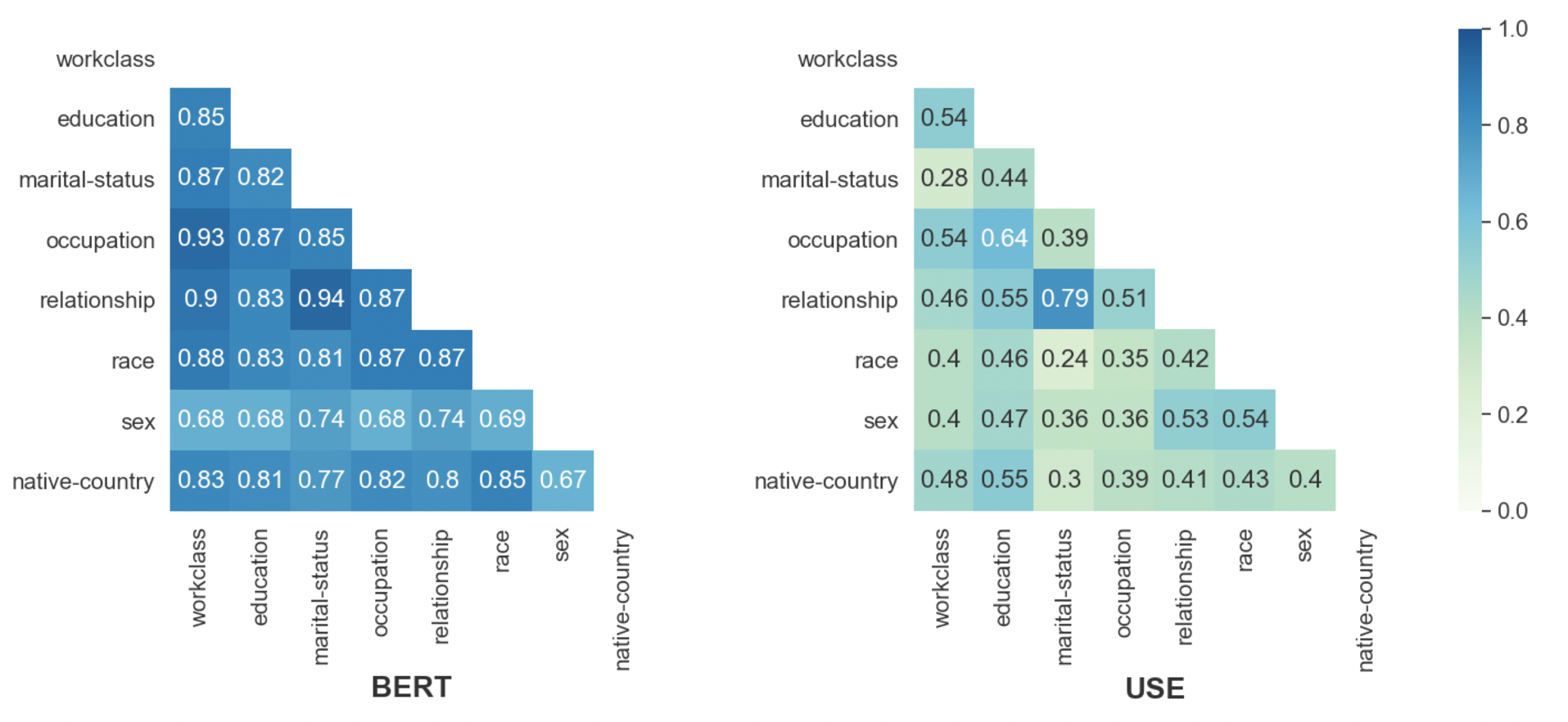}
    \vspace{-2.5ex}
    \caption{Experiment 9 --  Cosine similarities between the instance representations of the attributes of the Adult dataset for representations based on BERT and Google USE. For BERT, dissimilar attributes also have similarity values of at least 0.65, while for Google Use the range is much wider.}
    \label{fig:similarity_bert_use}
\end{figure*}

In Section~\ref{subsec:table_instance} we introduced the matching of attributes using instances and based on sampling.
Again, we use the same embedding for instances of a column as for the table based matcher
using either SUM or AVG to compute a vector representation for all (sampled) values of a column.
Based on such a column embedding, the cosine distance is used as similarity metric.

\paragraph*{Experiment 6: Robustness of Instance Representations}

For this experiment we again use the OAEI benchmark and additionally two film data sets from the Internet Movie Database (IMDB)\footnote{\url{https://github.com/AhmedSalahBasha/schema-matching/blob/master/imdb.csv}} and the Rotten Tomatoes Database (RT)\footnote{\url{https://github.com/AhmedSalahBasha/schema-matching/blob/master/rotten_tomatoes.csv}}.
In addition to 6000 to 7500 real film titles, these two data sets contain names of actors, genres and countries.
To test the quality of the representation of a column, subsets of the attribute instances are formed according to different patterns, represented and then compared to other representations of the same attribute.
We test the following patterns using this data sets:

\begin{description}
\item \textbf{Overlapping} Divide Column in two random but equally sized parts (might contain overlaps).
\item \textbf{Distinct} Remove duplicates and then split the remainder into two parts of equal size.
\item \textbf{N-Random} Take $N$ random distinct samples and split those into two parts.
\end{description}

Our results show that for all attributes, the cosine similarity is nearly one for the overlapping method, and at least $0.9$ for the distinct method. 
Even with $N=10$, hence only $5$ instance per vector, the representations have a similarity of at least $0.8$ (e.g., for IMDB genres or surnames).
If the instances have some internal structure (for example, non-normalized addresses consisting of streets and house numbers), the similarity measures are even closer to $1$ even for these very small samples.

However, it is not sufficient that the representations are robust even with sampling, they must also differ as much as possible from the representations of the other attributes.
Therefore, in the following we will further investigate the effect of sampling.

\paragraph*{Experiment 7: Effects of Instance Sampling}

For this experiment we again use the two movie data sets and additionally the movie-specific IIMB ontology from the OAEI benchmark.
For all attributes, representations without sampling and with Distinct N-Most-Common sampling (see Section~\ref{subsec:table_instance}) are calculated.
The sample size is adapted to the attribute in a way that it takes into account all instances appearing more frequently than average (considering the standard deviation).
The results using the widely used BERT embeddings can be found in Figure \ref{fig:sample_effect}.
For attributes with a very high degree of similarity, it usually remains high even after sampling, especially for attributes that are actually semantically similar (e.g., \texttt{IMDB\_title} and \texttt{RT\_title}).
Faulty matches are partially attenuated by sampling, but there are cases where the similarity remains high (e.g., \texttt{IIMB\_actor\_born\_in} and \texttt{RT\_title}).
Generally it is found that sampling is useful for spreading the range of values and thus ensuring that dissimilar attributes have a lower similarity value while the similarity value of related attributes remains high.

Now that we have examined the components of the instance-based approach, we will evaluate the quality of the actual matching and compare it to other approaches, both traditional and embedding-based.

\paragraph*{Experiment 8: Instance-Based Attribute Matching}
Analogous to Experiment 4, we again use the OAEI benchmark and compare our approach with a COMA reference implementation.
As baseline we use the more complex COMA 3.0 matching workflow \texttt{AllContextInstW}.
Among other things, it analyzes schema information such as table paths and element names, but also includes the instances themselves and, according to the authors, should (only) be used if instance information is available.
It is meant to be used stand-alone.
In order to avoid introducing additional parameters, we compare it to an embedding-based matcher that works purely on instances.
For productive use, however, it would make sense to combine this approach with a name-based approach and the correspondence pre-selection step.

The average results can again be found in Table \ref{tab:coma_use_avg}, the results for the individual matching problems can be found in Figure \ref{fig:coma_38}.
Our approach provides very high precision.
The seemingly low recall can be explained by the fact that there are instances for only about 1/3 of the attributes.
Most of the correspondences of these attributes can be found by the instance-based matcher, which can roughly maintain its quality-level even for the problems where the other three approaches have severe difficulties.
Instance-based matching can therefore help to find correspondences when attribute names are not meaningful or semantically difficult to compare, for example due to domain-specific language.
Overall, one can assume that a balanced combination of our two approaches would beat both the simpler and the complex (combined) baseline approach.

\subsection{How useful is Google USE?}

For the evaluation of our methods in this paper we mainly used vector representations based on Google USE (see Section \ref{subsec:embedding_approaches} for the advantages).
In the following we will briefly draw a comparison to the well-known BERT embeddings.

\paragraph*{Experiment 9: Quality of Word Embedding Model}
For this we proceed analogously to Nozaki et al. \cite{nozakietal}, who investigated instance-based matching with the classical Word Embedding model \textit{Word2Vec}:
For all attributes an averaged representation is calculated as described in Section \ref{subsec:table_instance} without sampling---once with BERT and once with Google USE.
This representation is then compared with the representation of all other attributes.

The results for this experiment on the Adult dataset,\!\footnote{\url{http://archive.ics.uci.edu/ml/datasets.html}} which Nozaki et al. used in their experiments too, can be found in Figure \ref{fig:similarity_bert_use}.
In both models, the similarity metric is highest between the attributes marital-status and relationship, which correspond semantically.
However, with BERT, even very different attributes have similarity metric values of at least $0.65$, while the range is significantly larger when using Google USE.
This shows that at least without any special fine-tuning the use of Google USE is preferable for the purpose of attribute matching.

\subsection{Runtime}

As a reference for the determination of the computation time for semantic representations with Google USE, we measured the runtimes for the RT dataset on a server with NVIDIA\textregistered{} TESLA\textregistered{} V100 graphics card with 16GB memory. 
The implementation of the instance-based and thus most complex matching took under three minutes to calculate all column vectors of the RT dataset (which has 10 columns with an average of $10,000$ instances).

\section{Conclusion \& Future Work}
\label{sec:conclusion}

We have shown that neural word embeddings can be utilized to propose a small set of possible candidates for schema matching which is crucial for data integration.
Our experiments prove that word embeddings can be used to bridge the semantic gap for several matching variants.
Our approach can be used instead of but in particular as a supplement to existing syntactic and semantic matchers.

The use of models pre-trained for general tasks seems to be sufficient as long as the database does not predominantly contain abbreviations and very specific terms.
It is advisable to use contextualized embeddings.

Both structural data like schema information and comments as well as the textual data instances themselves can be used for matching, whereby the effectiveness of the individual approaches strongly depends on the schemata.
In instance-based approaches, sampling can help to increase the distinction between similar and dissimilar attributes while reducing the number of instances to be considered.
This also makes it possible to use the approach for databases with very large numbers of instances.
Instance-based approaches work well with different types of entities. 
Weaknesses are found, for example, with attributes that all contain human names:
the embeddings are good for finding other attributes with names, but a further subdivision (e.g., by actors and directors) is difficult.

In the future, it would be interesting to further look at pre-processing of the textual elements before they are represented by embeddings (e.g., resolving of acronyms) and to include additional auxiliary information such as thesauri.
It should also be checked whether the results can be further improved by fine-tuning of the embeddings:
on the one hand, this could be done by additional training on corpora with domain-specific texts, on the other hand, analogous to Kolyvakis et al. \cite{kolyvakis2018deepalignment}, one could try to adapt the word vectors to the existing schema matching problem by using structured external domain knowledge.
As an orthogonal problem, the instance-based approach could be further improved to additionally support purely numerical attributes, i.e., to examine if it is possible to learn some kind of embedding for sets of numerical values beyond existing approaches like regular expressions, value range limits, averages, etc.
Finally, one could also further investigate matchers that use embeddings to compare natural language comments, which are available for some schemata, with each other or with table and attribute names.

\section*{Acknowledgments}
This work has been supported by the German Research Foundation as part of the Research Training Group \textit{Adaptive Preparation of Information from Heterogeneous Sources} (AIPHES) under grant No. GRK 1994/1.

\balance{}
\bibliographystyle{abbrv}
\bibliography{main}

\end{document}